\documentclass[aps,twocolumn,prl,reprint,amsmath,amssymb,showpacs,superscriptaddress]{revtex4-1}

\usepackage{multirow}
\usepackage{cancel}
\usepackage{graphicx,epsfig}
\usepackage{wasysym}
\usepackage{color,soul} 
\usepackage[normalem]{ulem}
\usepackage[colorlinks=true,citecolor=blue,linkcolor=red,linktocpage=true,pagebackref=false]{hyperref}
\def \beq {\begin{equation}}
\def \edq {\end{equation}}
\def \veps {\varepsilon}

\begin{document}
\title{
Probing the energy reactance with adiabatically driven quantum dots}

\author{Mar\'{\i}a Florencia Ludovico}
\affiliation{International Center for Advanced Studies, UNSAM, Campus Miguelete, 25 de Mayo y Francia, 1650 Buenos Aires, Argentina}
\affiliation{The Abdus Salam International Centre for Theoretical Physics, Strada Costiera 11, I-34151 Trieste, Italy} 
\author{Liliana Arrachea}
\affiliation{International Center for Advanced Studies, UNSAM, Campus Miguelete, 25 de Mayo y Francia, 1650 Buenos Aires, Argentina} 
\affiliation{Dahlem Center for Complex Quantum Systems and Fachbereich Physik, Freie Universit\"at Berlin, 14195 Berlin, Germany}
\author{Michael Moskalets}
\affiliation{Department of Metal and Semiconductor Physics,
NTU "Kharkiv Polytechnic Institute", 61002 Kharkiv, Ukraine}
\author{David S\'anchez}
\affiliation{Institute for Cross-Disciplinary Physics and Complex Systems IFISC (UIB-CSIC), E-07122 Palma de Mallorca, Spain}

\begin{abstract}
The tunneling Hamiltonian describes a particle transfer from one region to the other. 
While there is no particle storage in the tunneling region itself, it has associated certain amount of energy.
We name the corresponding flux {\em energy reactance} since, like an electrical reactance, it manifests itself in time-dependent transport only.  
Noticeably, this quantity is crucial to reproduce the universal charge relaxation resistance for a single-channel quantum capacitor at low temperatures.  
We  show that a conceptually simple experiment is capable of demonstrating the existence of   the energy reactance. 
\end{abstract}


\maketitle

{\em Motivation}. 
A very exciting experimental activity is lately taking place in search of controlling on-demand quantum coherent 
charge transport in the time domain. 
The recent burst of activity started with the experimental realization of quantum capacitors in quantum dots under ac driving \cite{Gabelli:2006eg}, single particle emitters \cite{Feve:2007jx}, and was followed by the generation of quantum charged solitons over the Fermi sea (levitons) \cite{Dubois:2013ul}. 
A controlled manipulation of flying single electrons \cite{Hermelin:2011du,McNeil:2011ex,Bertrand:2016ik} and their time-resolved detection \cite{Fletcher:2012te} have already been reported~\cite{Splettstoesser:2017jd}. 
These marvelous developments, along with the identically impressive progress in the field of fast thermometry \cite{zgi17,pek13,gas15}, are opening an avenue towards the study and control of the concomitant time-dependent energy flow in the quantum realm.

The relevant systems are characterized by small (nanoscale) components confining a small number of particles in contact to macroscopic reservoirs. This puts the description of the energy transport and heat generation beyond the 
scope of usual thermodynamical approaches, motivating a number of formal theoretical developments in statistical mechanics~\cite{Kosloff:2013cr} and condensed matter physics~\cite{Ludovico:2016hh}.  
At the heart of this problem, there is the proper definition of the quantum heat current in the time domain. The concept of heat looks very intuitive and anyone can provide a definition for it. 
Formally, it is a clear and well established concept in macroscopic systems close to equilibrium.
However, its accurate definition at the nanoscale and in situations away from equilibrium is a deep and subtle issue, in particular due to the coupling between a nanosystem and macroscopic reservoirs; see, e.g., Refs.~\onlinecite{Wu:2008gk,us1,Ankerhold:2014ky,rossello,Esposito:2014wu,Esposito:2015bg,Esposito:2015bg,Dare:2016jk,Bruch:2015ux,Ochoa:2016ee,us2}. 
In fact, while charge and energy are concepts obeying strict fundamental conservation laws, the definition of heat implies the proper identification of a portion of the total energy. 

 \begin{figure}[t]
  \includegraphics[width=0.4\textwidth]{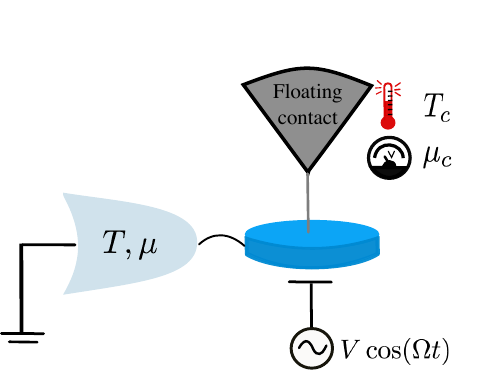}
  \caption{Schematic of our proposal. The quantum $RC$ circuit consists of a quantum dot (blue disk) coupled to a fermionic reservoir (light blue region) with
  well defined temperature $T$ and chemical potential $\mu$. Electrons can be transferred between the dot and the reservoir (black curve). The dot is capacitively
  coupled to a gate terminal where an ac potential of amplitude $V$ and frequency $\Omega$ is applied. A floating contact is also attached to the dot. The temperature $T_c$ and chemical potential $\mu_c$ of the floating contact adjust themselves to cancel both the charge and heat current flowing through it, thus allowing for an experimental test of the energy reactance, namely, the variation of the stored energy at the tunneling region between the floating contact and the dot (gray line).}\label{fig1}
\end{figure}
 
An appealing scenario to address this problem from the theoretical point of view is a periodically driven single level in contact to an electron reservoir. 
This is the most basic and meaningful setup to  analyze the interplay of charge and energy dynamics. At the same time, 
 this is the simplest model 
for a quantum $RC$ circuit \cite{Buttiker:1993wh}, which  has been experimentally realized \cite{Gabelli:2006eg}. A sketch is presented in Fig.~\ref{fig1}, where we stress that the driven level represents a quantum dot.
The nonequilibrium ingredient is provided by the time-dependent gate voltage $V(t)= V \cos\left( \Omega t \right)$ locally applied to the single level. The reservoir is an electron gas with temperature $T$ and chemical potential $ \mu$, and the strength of the coupling  between the two subsystems is arbitrary. The setup also includes a ``floating contact'', which we will discuss in detail later on.

The effect of the periodic driving is twofold. On one hand, it induces a charge current that flows
between the dot and the reservoir as a function of time. On the other hand, it performs a work on the system, thus injecting energy that is ultimately dissipated as heat deep inside the reservoir.  
Importantly, due to charge conservation, the electronic current is defined as the change in time of the electron number either at the reservoir or on the dot. No  contribution of the tunneling region on the charge current exists.  
In contrast, the energy delivered by the external ac source is temporarily stored in three different parts of the setup: the dot, the reservoir and also  in the dot-reservoir tunneling region. The role of the tunneling region is typically disregarded in classical thermodynamics because it is a surface term that is negligible
when both the system and the reservoir are macroscopic~\cite{bellac}.
Yet, in the nanoscale setup studied here the amount of energy stored in
the dot is comparable to that of the tunneling region and the latter can no longer be neglected. 

In a recent work \cite{us1} we have coined the name of {\em energy reactance} to characterize the energy temporarily stored at the tunneling region. 
This is a thermal analogue of an electrical reactance (due to  electrical capacitance and inductance), which manifests itself in a time-dependent setup only.  
We have argued that it is physically meaningful to take the energy reactance into account as a contribution to the time-dependent heat current flowing into the reservoir. 
We have shown that this is in full agreement with the laws of thermodynamics~\cite{us2,Ludovico:2016hh}.  
While some recent works raised some concerns~\cite{Esposito:2014wu,Ochoa:2016ee}, other works arrived at conclusions similar to our analysis~\cite{Wu:2008gk,rossello,Bruch:2015ux}.  
The aim of the present work is to make one step further by proposing a measurement scheme that is able to test the effect of the energy reactance  onto a time-dependent heat flux. 

\textit{Proposed experiment and predictions.} The setup is sketched in Fig.~\ref{fig1}, where we introduce a floating contact attached (e.g., via tunneling) to the quantum dot. 
When a periodic gate voltage $V(t)$ is applied, charge and heat currents enter not only the reservoir but also the floating contact. The latter can adjust its chemical potential $\mu_{c}$ and temperature $T_{c}$ to maintain zero charge and 
heat currents flowing into it.
We will focus on slow ``adiabatic'' driving, which corresponds to a driving period  much larger than any characteristic time scale for the system. 
Assuming that the charge and energy relaxation rate of the floating contact is much faster than any other characteristic time, $\mu_{c}$ and $T_{c}$ will change
instantaneously to prevent charge and heat accumulation on the floating contact.
In contrast, the reservoir is a massive electrode that keeps its temperature
and chemical potential constant independently of the ac potential.
In practice, this can be achieved grounding the reservoir as indicated in Fig.~\ref{fig1}. Its temperature variations would
be suppressed if the reservoir has in addition a large heat capacity.

The evolution of the chemical potential and temperature of the floating contact as the dot is aidabatically driven can be sensed by means of a voltage probe and a thermometer~\cite{Engquist:1981wz,dubi,benenti,sanchez,bergfield,Battista:2013ew,thermo,probe,ye}, as indicated in the figure.
We predict different behaviors for $\mu_{c}$ and $T_{c}$ depending on whether the energy reactance is considered or not in the heat flux into the floating contact. In this way, the proposed experiment would help to discern
on the proper definition of the heat current and test the existence of the energy reactance. 

The  results are the following:
(i) By defining the heat flux into the
floating contact, taking into account the energy reactance as in Ref.~\onlinecite{us1}, the temperature of the floating contact does not vary with time. The outcome is 
\begin{eqnarray}
T_{c} = T \,,
\label{main}
\end{eqnarray}
where  $T$
 is the background temperature. 
The chemical potential of the contact $ \mu_{c}(t)$ does vary  with time in a periodic fashion with a period dictated by the electrical current flowing through the dot. 
(ii) We demonstrate that any other definition of the heat current,  that does not properly account for the energy reactance, necessarily leads to a change in both quantities, $T_{c}(t)$ and
$\mu_{c}(t)$ as functions of time.

{\em Heat current into the floating contact and quantum energy reactance.}
Let the rates of change  for the charge  and 
the internal energy stored in the floating contact 
 due to exchanges with the rest of the device be, respectively, 
 $\dot{N}_c(t) $ and $\dot{U}_c(t)$. Similarly, the rate of change for the
 energy stored at the tunneling region between the dot and the floating contact is denoted by
 $\dot{U}_{T_c}(t)$. 
The meaningful definition for the instantaneous heat current entering the floating contact is~\cite{us1}
\beq\label{heat}
\dot{Q}_c(t)=\dot{U}_c(t)+\frac{\dot{U}_{T_c}(t)}{2}-\mu_c(t)\dot{N}_c(t)\,.
\edq
The energy reactance, $\dot{U}_{T_c}(t)/2$, contributes to the heat flux only instantaneously and as such vanishes when averaged over one driving period.

From the theoretical point of view, the energy reactance is necessary to 
derive an instantaneous Joule law 
for the heat  current into a (single-channel) floating contact at low temperatures,
$\dot{Q}_c(t)= R_q [\dot{N}_c(t)]^2$,
with the universal charge relaxation resistance, the B\"{u}ttiker resistance,  $R_{q}= h/2e^2$ \cite{Gabelli:2006eg,Buttiker:1993wh}.
The energy reactance is also necessary to both
reconcile  the relation between the Green function and  the scattering matrix formalisms \cite{Arrachea-floquet} for the instantaneous heat current \cite{us1} and also to obtain correct frequency parity properties of the response functions \cite{rossello}. 

{\em Temperature and chemical potential of the floating contact.}
Our goal is to explicitly show that the definition of Eq.~(\ref{heat}) can be verified by measuring the temperature and chemical potential of the floating contact.  The latter is a small conductor that we assume to be in thermal equilibrium
at every instant of time such that both temperature and chemical potential adjust themselves to satisfy the condition of vanishing charge and vanishing heat current, i.e., $\mu_c(t)$ and $T_c(t)$ simultaneously fulfill  $\dot{Q}_c(t)=\dot{N}_c(t)=0$.
The local equilibrium condition is justified
in the adiabatic regime (very low driving frequency $\Omega$),
mostly accessible in experiments~\cite{Gabelli:2006eg}.
Deviation of the floating contact temperature and chemical potential from their
stationary values are denoted by $\delta T_c(t)=T_c(t)-T$ and 
$\delta \mu_c(t)= \mu_c(t)-\mu$, respectively.
In the adiabatic regime, these quantities are small,  $\delta T_c(t) ,\delta \mu_c(t) \propto \hbar\Omega$. 
As a consequence, we can evaluate both charge and heat fluxes in linear response in these quantities (while the amplitude of the ac driving potential is arbitrary). 

Following Refs.~\onlinecite{Ludovico:2015uw,us2} we  expand the fluxes ${\bf J} (t)\equiv \left(\dot{N}_c(t),\;\dot{Q}_c(t)\right)$  in the affinities 
 ${\bf X} =\left( \delta \mu_c(t), \; \delta T_c(t), \;\hbar \Omega \right)$, with coefficients  $\Lambda_{ij}(t)$ as
\beq\label{linearresp}
J_i(t)=\sum_{j=1}^{3}\Lambda_{ij}(t) X_j (t),
\edq
where  $i=1,2$ ($j=1,2,3$)  label the different components of the vectors
${\bf J}$ and ${\bf X}$, respectively. The coefficients of the above expansion are response functions evaluated with the frozen Hamiltonian at time $t$ and have the following
physical interpretation: $\Lambda_{11}$ and $\Lambda_{22}$ are the usual electric and thermal conductances. On the other hand, $\Lambda_{12}$ 
(related to the Seebeck effect) 
and $\Lambda_{21}$ (related to the Peltier effect)
capture the thermoelectric transport, and they satisfy the reciprocity relation $\Lambda_{21}=T\Lambda_{12}$~\cite{onsager,buttiker,matthews,revcasati}. 
Finally, ${\Lambda}_{13}$ and ${\Lambda}_{23}$ 
describe, respectively, 
the generation of charge and heat currents by the ac driving. They also obey Onsager relations with the coefficients entering the work flux (not considered here) \cite{Ludovico:2015uw}.
Explicit expressions of these coefficients will be supplied below for the specific model.

Here, we notice that the conditions of vanishing fluxes to the floating contact amounts to 
finding the solution of the $2 \times 2$ linear set of equations 
$\sum_{j=1}^2 \Lambda_{ij}X_j =- \Lambda_{i3} \; \hbar \Omega,\;i=1,2$.
The solutions are
\begin{eqnarray} 
\delta \mu_c(t) & = & \frac{\Lambda_{12} \Lambda_{23}- \Lambda_{13} \Lambda_{22}}{\mbox{det}\,{\bf \Lambda}^{\prime}}\; \hbar \Omega ,\nonumber\\
\label{solset} \\
\delta T_c(t) & = & \frac{\Lambda_{13} \Lambda_{21}- \Lambda_{11} \Lambda_{23}}{\mbox{det}\,{\bf \Lambda}^{\prime}} \;\hbar \Omega,
\nonumber
\end{eqnarray}
where ${\mbox{det}\,{\bf \Lambda}^{\prime}}$ corresponds to the determinant of the $2\times 2$ matrix determined by the condition $j\neq 3$.

The coefficients $\Lambda$ can be calculated for the system considered
in Fig.~\ref{fig1} following Refs. \onlinecite{Arrachea-floquet, dyson} (see details in Ref. \onlinecite{sm})
\beq\label{coeff1}
\Lambda_{ij}(t) =
  \begin{cases}
    \int   \frac{(\veps-\mu)^{i+j-2}}{hT^{(j-1)}}\mathcal{T}(t,\veps)\, \partial_\veps  f \,d\veps  &  \text{if } j\neq 3 \\
    \\
     -\frac{\Gamma_c\dot{V}}{(\Gamma+\Gamma_c) h\Omega }\int  (\veps-\mu)^{i-1}
     \rho_f(t,\veps) \,\partial_\veps f \,d\veps   &  \text{if } j=3\,,
\end{cases}
\edq
The distinction between $j\neq 3$ and $j=3$ is important.
In the former case, the response depends on the instantaneous transmission probability $\mathcal{T}(t,\veps)$ for electrons traversing the quantum dot between the reservoir and the floating contact. Physically, this corresponds to dc transport.
In the latter case, the response is a function of the time derivative of the potential applied to the gate, $\dot V = - \Omega V \sin\left( \Omega t \right)$,
and the instantaneous local density of states of the dot, $\rho_f(t,\veps)$.
Physically, this is pumping and, as such, of ac nature.
Both coefficients are time dependent because the system adiabatically
reacts to the instantaneous ac driving potential~\cite{mos12}.
Finally, in Eq.~\eqref{coeff1}
$f$ is the Fermi-Dirac distribution of the reservoir,
while $\Gamma_c=|w_c|^2 \rho_c$ and $\Gamma=|w|^2 \rho$ are the hybridization functions with $w_c$ the dot-floating contact couplings and $w$ the dot-reservoir couplings. The density of states of the floating contact is $\rho_c$ and that of the reservoir is $\rho$.

Interestingly, we readily find that the coefficients of Eq.~(\ref{coeff1}) satisfy {\it i}) $\Lambda_{13} \Lambda_{21}- \Lambda_{11} \Lambda_{23}=0$ and {\it ii}) $\Lambda_{j3}=-\Lambda_{j1}\frac{\dot{V}}{\Gamma\Omega}$ with $j=1,2$, leading to the solution 
 \begin{equation}\label{sol}
 \delta T_c(t)=0,\;\;\;\;\;\;\;\;\;\;\;\;\;\;\;\;\;\;\;\;\;
\delta\mu_c(t) = \frac{\hbar}{\Gamma}e \dot{V}\,.
\end{equation}
From Eq.~(\ref{solset}) and the relation {\it ii}), we see that the above results do not actually depend on coupling to the floating contact.
We have checked that Eq.~\eqref{sol} is valid for any temperature $T$, provided that
 the adiabaticity condition  $\Gamma, \Gamma_c \gg \hbar\Omega$ is satisfied \cite{brouwer,spletti08}. This is true
even for temperatures close to zero, in which case the second order contributions in the affinities should be added to Eq. (\ref{linearresp}) \cite{sm}.

In summary, the floating contact fulfills the conditions of vanishing heat and charge fluxes by changing $\delta \mu_c(t)$ in time according to Eq.~(\ref{sol}) while keeping its temperature constant and equal to the background temperature, as indicated in  Eq. (\ref{main}).

\begin{figure}[b]
  \includegraphics[width=0.52\textwidth]{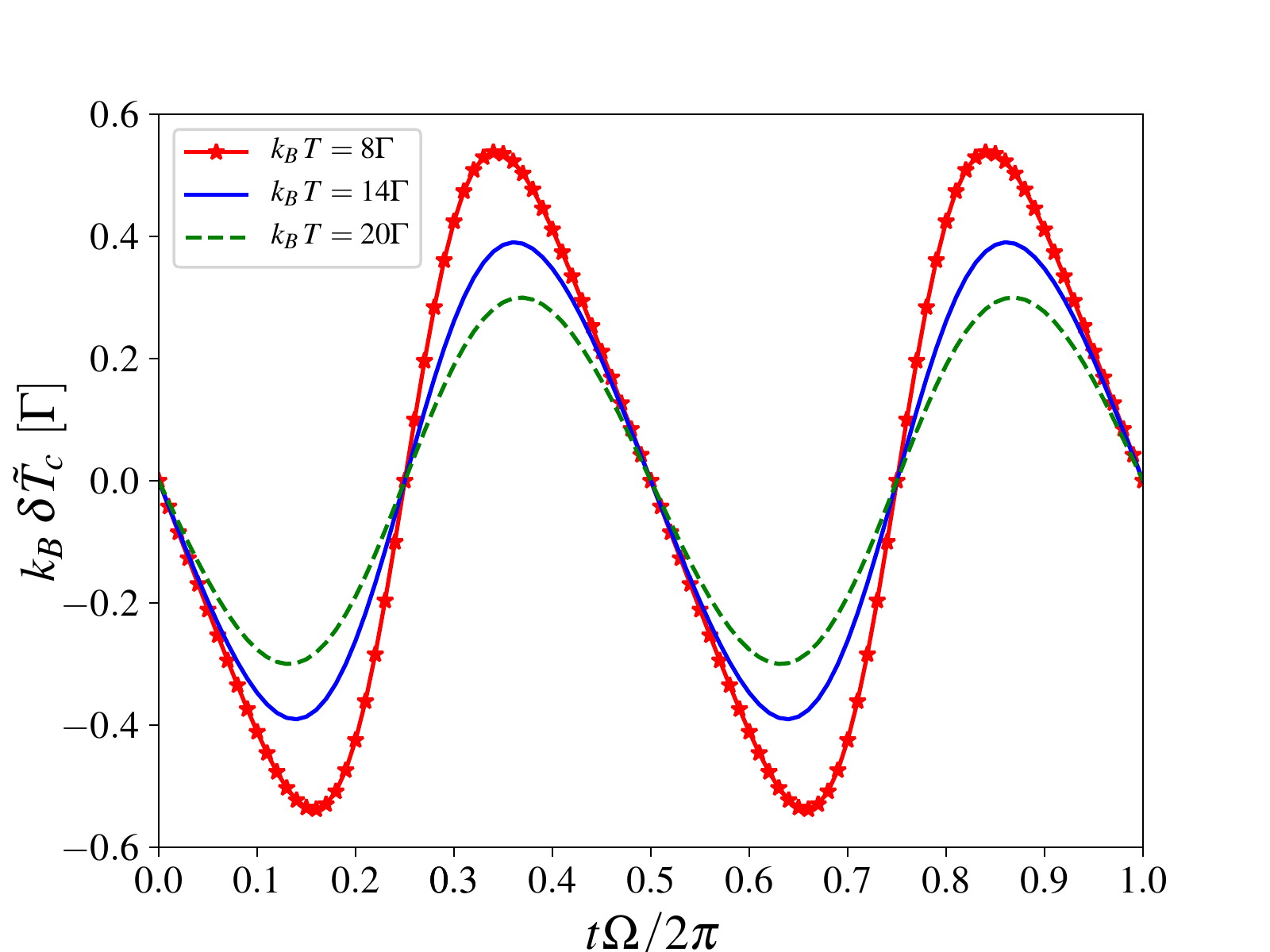}
  \caption{Deviation of the temperature of the floating contact, $\delta\tilde{T}_c$, as a function of time for different  background temperatures $T$. The ac potential is $20\,\Gamma\cos(\Omega t)$ with $\hbar\Omega=0.07\Gamma$. The hybridization between the floating contact and the quantum level is $\Gamma_c=0.6\Gamma$. All energies are expressed in units of the hybridization $\Gamma$ with the reservoir. The temperature of the floating contact displays oscillations that depend on the background temperature. As $T$ increases, the oscillations become less pronounced and the maxima positions deviate from the moment when the level is aligned with the chemical potential of the reservoir, which in this case corresponds to $t\Omega/2\pi=0.25$ and $t\Omega/2\pi=0.75$.}
\label{fig2}
\end{figure}

{\em Examine the energy reactance}.
We would like to stress now that Eq.~(\ref{sol}), in particular, the prediction of a constant temperature of the floating contact expressed in Eq. (\ref{main}), constitutes a proof for the existence of the energy reactance $\dot{U}_{T_c}(t)/2$ and the definition of the heat current  as in Eq.~(\ref{heat}). 
This can be easily understood by noticing that we would arrive at completely different conclusions on the behavior of the temperature of the floating contact if we consider a definition of the heat flux that does not take into account the energy reactance. 

As a proof, let us analyze the consequence of adopting  a commonly used definition, that does not take into account the energy reactance. This corresponds to the following expression for the heat flux into the floating contact,
\beq\label{heatilde}
\dot{\tilde{Q}}_c(t)=\dot{U}_c(t)-\mu_c(t)\dot{N}_c(t).
\edq
We need to recalculate the coefficients $\Lambda_{2j}(t)$ by using the above equation.  
We denote the so defined coefficients by $\tilde{\Lambda}_{2,j}(t)$. 
From Eq.~(\ref{solset}), where we replace $\Lambda_{2,j}(t) \rightarrow \tilde{\Lambda}_{2,j}(t),\; j=1,2,3$, we find the floating contact temperature $\delta\tilde{T}_c(t)$ and chemical potential $\delta \tilde{\mu}_c(t)$.
In contrast to Eq.~(\ref{sol}), now we find that both the temperature $\delta\tilde{T}_c(t)$ and the chemical potential $\delta\tilde{\mu}_c(t)$ of the floating contact change in time. In the case of the chemical potential, $\delta\tilde{\mu}_c$ evolves in time in a different pattern from that described by Eq.~(\ref{sol}). We turn to focus on the behavior of the temperature $\delta\tilde{T}_c$, which is shown in Fig.~\ref{fig2}.
It is worth noting that the amplitude of the
$\delta\tilde{T}_c$ oscillations decreases as $T$ increases,
which shows that the two definitions of the heat current agree in the high temperature limit. These results show that  the role of the energy reactance is particularly relevant in the quantum regime. In the classical high-temperature limit the temperature 
 of the floating contact is independent of time, either with the heat current defined  as in Eq, (\ref{heat}) or with the definition of Eq. (\ref{heatilde}).

{\em Conclusion.} We have shown that the behavior of the time-resolved chemical potential and temperature of a floating contact coupled to an adiabatically driven quantum dot is strongly sensitive on the definition of the instantaneous heat flux. 
For this reason, sensing these quantities would provide an experimental test for the relevance of the energy reactance introduced in
Ref.~\cite{us1} as a component of the time-dependent heat flux. 

Specifically, for an adiabatically driven quantum dot with a single active level coupled to a single reservoir, we have shown that: 
(i) If the energy reactance is taken into account, then the temperature of the floating contact is constant and equal to that of the reservoir, while its chemical potential follows  the time derivative of the driving potential, $\dot{V}$, as expressed in Eq.~(\ref{sol}). 
Instead, (ii) if the energy reactance is not taken into account, these two quantities follow a nonuniversal and rather cumbersome time-dependent pattern.

The experiment we propose is close to  the scope of present-day experimental techniques.  In fact, typical level spacing for quantum dots is around $100 $~$\mu$eV~\cite{kouw}. Thus, by keeping driving amplitudes below this energy, we would basically have a single active level. On the other hand, typical parameters for single particle emitters  have $\Gamma \simeq 1 \mu$eV ($\simeq 1$~GHz) and are operated at frequencies 
$\Omega \simeq 0.1$~GHz \cite{Gabelli:2006eg}, which satisfy the adiabatic condition $\hbar \Omega  < \Gamma $. 
As a consequence, a fast thermometer~\cite{zgi17} is able to follow temperature changes of the floating contact on the nanosecond scale. 
Experiments are typically performed at temperatures close to $T \sim 100$~mK. For this temperature, the oscillations in the temperature shown in Fig.~\ref{fig2} have an amplitude of $\delta \tilde{T}_c \simeq  10$~mK.

We emphasize that the question about the role of the energy reactance in the definition of a time-dependent heat flux is a fundamental one. 
It is not restricted to slowly driven systems of noninteracting electrons but is also relevant for interacting models, for fast drivings, and for weakly and strongly coupled systems. So far this question has been addressed  only theoretically. 
The present proposal shows that a thermometer probe response will experimentally demonstrate the existence of the energy reactance.

{\em Acknowlegments}. This work was supported by MINECO under Grant No.~FIS2014-52564, UBACyT, CONICET and MINCyT, Argentina. LA thanks the support of the Alexander von Humboldt Foundation.
MM thanks the support and the hospitality of the Aalto University, Finland.

\newpage
\setcounter{figure}{0}
\setcounter{equation}{0}
\renewcommand{\thefigure}{S\arabic{figure}}  
\renewcommand{\theequation}{S\arabic{equation}} 
\renewcommand{\figurename}{Figure} 
\renewcommand{\thefootnote}{S\arabic{reference}}

\pagebreak
\onecolumngrid
\pagebreak

\section*{Probing the energy reactance with adiabatically driven quantum dots: Supplementary Information}

This supporting document describes in detail the derivation of the time-dependent linear response coefficients for particle and heat currents flowing into the floating contact.

\subsection{Theoretical model}

We consider a simple setup, with all the necessary ingredients to analyze the dynamical energy transfer and to probe the energy reactance. It is the most basic model for a quantum capacitor, which consists
in a periodically driven single level (quantum dot) coupled to an electron bath, the reservoir,  
with temperature $T_r=T$ and chemical potential $\mu_r=\mu$. The time-dependent driving is 
provided by the application of an oscillatory gate voltage of the form $V(t)=V{\cos}(\Omega t)$. 

To test the effect of the energy reactance on the time-dependent heat flux, we introduce a floating contact, 
which is coupled to the driven level. In order to be electrically and thermally isolated from the environment at every time, the floating contact
instantaneously adjusts its chemical potential $\mu_c$ and temperature $T_c$. 

The Hamiltonian of the full system, the quantum capacitor together with the floating contact,  can be separated into three contributions,
\beq
H(t)=H_{QC}(t)+H_c+H_{T_c}.
\edq
The Hamiltonian $H_{QC}$ represents the quantum capacitor, which contains three elements: the single level, an electron bath 
(the reservoir, denoted by the letter $r$),    
and the coupling between the two. 
Then,
\beq
H_{QC}(t)=\sum_{k}\left[\veps^r_{k} c^\dagger_{k} c_{k}+w (d^\dagger c_k+ c_k^\dagger d)\right]+(\veps_0+V(t))d^\dagger d,
\edq
where $\veps_{k}^r$ is the energy band of the 
reservoir and $w$ is the coupling amplitude to the driven level. The energy $\veps_0$ corresponds to the bare level, which for simplicity will be considered aligned with the chemical potential of the 
reservoir, i.e. $\veps_0=\mu$ .
The operator $c^\dagger_{k} (c_{k})$ create (destroy) an electron with a wavevector ${k}$ in the reservoir, while $d^\dagger$ and $d$ are associated to the degrees of 
freedom of the single level.  

Similarly, the floating contact is represented by the Hamiltonian,
\beq
H_c=\sum_{q}\veps^c_{q} a^\dagger_{q} a_{q},
\edq
while in this case, the operators $a_q$ and $a_q^\dagger$ are responsible, respectively, for the creation and destruction of an electron in the floating contact with an energy $\veps^c_{q}$. The coupling between the level and the floating contact can be written as
\beq
H_{T_c}=\sum_{q} w^c (d^\dagger a_q+ a_q^\dagger d).
\edq

\subsubsection{Particle and heat currents}
Now, the aim is to compute the particle and heat fluxes entering the floating contact. The time variation of the particles present in the floating contact is given by the exact expression,
$\dot{N}_c=\frac{i}{\hbar}\langle [H, N_c]\rangle$.
To compute the time-dependent heat current as in Eq. (\ref{heat}), or by adopting the definition (\ref{heatilde}), we define the internal energy stored in the floating contact and in the tunneling region as $\dot{U}_\beta=\frac{i}{\hbar}\langle [H, H_\beta]\rangle$, 
with $\beta=c, T_c$. 

In Refs. \cite{us:prb} and \cite{Ludovico:entropy}, we presented that the evolution of the expectation value of an observable (e.g the number of particles or the energies) can be obtained by recourse to Keldysh non-equilibrium Green's functions. In this way, the different currents can be computed in terms of the retarded Green function $G^R(t,t')=-i\theta (t-t')\langle \{ d(t), d^\dagger (t')\}\rangle$ of the single level. For a time-periodic driving it is convenient to use the Floquet-Fourier representation \cite{Arrachea-floquet-1},
\beq
G^R(t,t')=\sum_{n=-\infty}^{\infty}\int^\infty_{-\infty} \frac{d\veps}{2\pi}e^{-i\frac{\veps}{\hbar}(t-t')}e^{-in\Omega t}{\cal G}(n,\veps).
\edq

As we showed in detail in Ref. \cite{us:prb} for a single driven level connected to many reservoirs, the particle current entering one of them, as for example the floating contact, is
\beq\label{chargesm}
\dot{N}_c(t)=\sum_l \int \frac{d\veps}{h} e^{-il\Omega t}\Gamma_c \left\{i{\cal G}^*(-l,\veps)\left[ f_c(\veps)-f_c(\veps_l)\right]-\sum_n\sum_{\alpha=r,c}\left[f_c(\veps)-f_\alpha(\veps_n)\right]\Gamma_\alpha {\cal G}(l+n,\veps_n){\cal G}^*(n,\veps_n)\right\},
\edq
where $\alpha=r$ corresponds to the reservoir of the capacitor, and $\alpha=c$ is the floating contact.  Some of the energies are shifted by a multiple of the energy quantum $\hbar \Omega$ as $\veps_n=\veps-n\hbar\Omega$. We have introduced the Fermi-Dirac distribution of the reservoir labeled by $\alpha$, $f_\alpha(\veps)=[e^{(\veps-\mu_\alpha)/{k_B T_\alpha}}+1]^{-1}$, with $k_B$ being the Boltzmann constant. The hybridizations with the reservoirs are $\Gamma_c=\vert w^c\vert^2 \rho_c$ for the floating contact and $\Gamma=\vert w\vert^2 \rho$ for the 
reservoir, with
$\rho_c=\sum_{k\in c}2\pi \delta(\veps-\veps_k^c)$ and $\rho=\sum_{k\in m}2\pi \delta(\veps-\veps_k^m)$ being the density of states of the floating contact and the 
reservoir, respectively. We are considering the wide band limit, in which $\Gamma_c$ and $\Gamma$ are constant functions.

In the same work \cite{us:prb}, and also in Ref. \cite{Ludovico:entropy} for a more general setup,  we showed that 
according to the definition in Eq. (\ref{heat}) the heat flux is
\begin{eqnarray}\label{heat1sm}
\dot{Q}_c(t) &=&\sum_l \int \frac{d\veps}{h} e^{-il\Omega t}\Gamma_c \Bigg\{i{\cal G}^*(-l,\veps)(\veps_{\frac{l}{2}}-\mu_c)\left[ f_c(\veps)-f_c(\veps_l)\right]\nonumber\\
&& \left.-\sum_n\sum_{\alpha=r,c}(\veps_{-\frac{l}{2}}-\mu_c)\left[f_c(\veps)-f_\alpha(\veps_n)\right]\Gamma_\alpha {\cal G}(l+n,\veps_n){\cal G}^*(n,\veps_n)\right\}.
\end{eqnarray}

If we adopt the definition in Eq. (\ref{heatilde}), which does not take into account the energy reactance $\dot{U}_{T_c}(t)/2$, then
\begin{eqnarray}\label{heat2sm}
\dot{\tilde{Q}}_c(t) &=& \sum_l \int \frac{d\veps}{h} e^{-il\Omega t}\Gamma_c \Bigg\{i{\cal G}^*(-l,\veps)\bigg[ (\veps-\mu_c)( f_c(\veps)- f_c(\veps_l))+l\hbar\Omega f_c(\veps_l) \bigg]\nonumber\\
& &-\sum_n\sum_{\alpha=r,c}\bigg[(\veps-\mu_c) (f_c(\veps)-f_\alpha(\veps_n))-\frac{l}{2}\hbar \Omega f_\alpha(\veps_n)\bigg]\Gamma_\alpha {\cal G}(l+n,\veps_n){\cal G}^*(n,\veps_n)\Bigg\}.
\end{eqnarray}

Here, we stress that all the above expressions for both particle and heat fluxes are exact, in the sense that they are valid for arbitrary values of the driving frequency, amplitude and temperature.   

\subsection{Linear response coefficients ${\bf\Lambda}$ and ${\bf{\tilde\Lambda}}$}

In what follows, we focus on the adiabatic regime, in which the driving frequency $\Omega$ is very low.  As we presented in our previous works \cite{us:prb, Ludovico:entropy}, to which we refer the reader for further details, we can expand the Floquet components ${\cal G}(n,\veps)$ up to first order in $\Omega$ as

\beq\label{lowfreq}
{\cal G}(n,\veps)\sim \int^\tau_0\frac{dt}{\tau}e^{in\Omega t}\left[G^f(t,\veps)+\frac{i\hbar}{2}\frac{\partial^2}{\partial t\partial\veps}G^f(t,\veps)\right] .
\edq

Here $\tau=2\pi/\Omega$ is the driving period, and 
\beq\label{frozeng}
G^f(t,\veps)=\left(\veps-\veps_0-V(t)-i\frac{(\Gamma +\Gamma_c)}{2}\right)^{-1}
\edq
is the frozen Green's function, 
which corresponds to the equilibrium solution of the Dyson equation \cite{dyson-1} at a given {\it frozen} time $t$.

Within the adiabatic regime, the departures of the temperature and chemical potential of the floating contact from those of the 
reservoir, $\delta T_c$ and $\delta\mu_c$, are proportional
to $\hbar\Omega$. Hence, we can also evaluate Eqs. (\ref{chargesm}), (\ref{heat1sm}) and (\ref{heat2sm}) within linear response in these quantities by expanding 
\beq\label{fermi1}
f_c(\veps_n)\sim f(\veps)-\partial_\veps f\, n\hbar\Omega-\partial_\veps f\,\delta\mu_c-\frac{(\veps-\mu)}{T}\partial_\veps f\,\delta T_c
\edq
where $f_r(\veps)=f(\veps)$ is the Fermi distribution of the 
reservoir.

Then, by using the expansions for slow driving (\ref{lowfreq}) and (\ref{fermi1}) in the expressions of the charge and heat fluxes, we can compute the linear response coefficients
${\bf \Lambda}$ as defined in Eq. (\ref{linearresp}) of the main text

\beq\tag{5}
\Lambda_{ij}(t) =
  \begin{cases}
    \int   \frac{(\veps-\mu)^{i+j-2}}{hT^{(j-1)}}\mathcal{T}(t,\veps)\, \partial_\veps  f \,d\veps  &  \text{if } j\neq 3 \\
    \\
     -\frac{\Gamma_c\dot{V}}{(\Gamma+\Gamma_c) h\Omega }\int  (\veps-\mu)^{i-1}
     \rho_f(t,\veps) \,\partial_\veps f \,d\veps   &  \text{if } j=3\,,
\end{cases}
\edq
where $\rho_f(t,\veps)=\vert G^f(t,\veps)\vert^2(\Gamma+\Gamma_c)$ is the total {frozen} density of states of the quantum dot, and $\mathcal{T}(t,\veps)=\vert G^f(t,\veps)\vert^2\Gamma_c\Gamma$ is the transmission probability.

The same procedure can be applied to the heat flux in Eq. (\ref{heat2sm}), whose definition does not take into account the energy stored in the tunneling region. In this case, we replace 
$\Lambda^{2j}\rightarrow \tilde{\Lambda}^{2j}$, so that
\beq
\dot{\tilde{Q}}_c(t)=\sum_{j=1}^3\tilde{\Lambda}_{2j}(t)X_j(t).
\edq
For this different definition of the heat, we find that $\tilde{\Lambda}_{21}(t)=\Lambda_{21}(t)$ and $\tilde{\Lambda}_{22}(t)=\Lambda_{22}(t)$, while the coefficient describing the pumping of heat changes as
\beq\label{crl}
\tilde{\Lambda}_{23}(t)= -\frac{\Gamma_c\dot{V}V}{(\Gamma
+\Gamma_c)h\Omega}\int{d\veps}\frac{df}{d\veps}\rho_f(t,\veps).
\edq

\subsection{Chemical potential and temperature of the floating contact in the zero temperature limit}

In the very low temperature limit of the 
reservoir, when $T\rightarrow 0$, an analysis of the fluxes within linear response
turns out not to be appropriate anymore since second order contributions could become dominant. In this case, instead of Eq. (\ref{linearresp}), the fluxes should be expanded as 
\beq
J_i=\sum_{j=1}^{3}\bigg(\Lambda_{ij}X_j+\sum_{m\leq j}L^i_{mj}X_mX_j\bigg),
\edq
where $\overrightarrow{{\bf L}}$ is a vector composed by matrices which capture the second order terms. The extreme situation occurs at $T=0$, in which absolutely all the linear response coefficients of the heat $\dot{Q}_c$ vanish ($\Lambda_{2j}=0$). In what follows, we focus on that case and compute both heat and particle fluxes entering the floating contact, in order to study if Eq. (\ref{sol}) remain valid when the temperature of the 
reservoir is close to zero. For the heat flux we find
\beq
\dot{Q}_c^{T=0}(t)=L^2_{22}(t)\delta T_c(t)^2+L_{13}^2(t)\delta \mu_c(t)\hbar\Omega+L_{33}^2(t)(\hbar\Omega)^2,
\edq
where 
\begin{eqnarray}\label{cheatto}
L^2_{22}(t)& = & -\frac{\pi^2}{3h}\mathcal{T}(t,\mu)\nonumber\\
L^2_{13}(t)& = & -\frac{\Gamma_c\Gamma\;\dot{V}}{(\Gamma+\Gamma_c) h\Omega}\frac{\rho_f(t,\mu)^2}{2},\\
L^2_{33}(t)& = & -\frac{\dot{V}}{\Gamma \Omega}L^2_{13}(t)\nonumber
\end{eqnarray}
and all other coefficients are zero,  $L_{11}^2=L_{12}^2=L_{23}^2=0$. However for the particle current, unlike the heat flux, only the coefficient $\Lambda_{12}=0$ at zero temperature while first order contributions in $\delta \mu_c$ and $\hbar\Omega$ remain. Thus, we can express the particle current at lowest order in the affinities as
\beq
\dot{N}_c^{T=0}(t)=\Lambda_{11}(t)\delta \mu_c(t)+\Lambda_{13}(t)\hbar\Omega+L_{22}^1(t)\delta T_c(t)^2,
\edq
with 
\begin{eqnarray}\label{cchargeto}
L^1_{22}(t)& = & -\frac{\pi^2}{3h}\partial_\veps\mathcal{T}(t,\mu).
\end{eqnarray}
Here, it is worth to mention that $L_{12}^1=L_{23}^1=0$, so that $L_{22}^1\delta T_c^2$ is the only lowest order contribution in $\delta T_c$.

Similarly, the chemical potential and the temperature of the floating contact can also be found by the condition $\dot{N}_c^{T=0}=\dot{Q}_c^{T=0}=0$. In this occasion, the vanishing fluxes condition leads to a set of equation which is linear en $\delta T_c^2$ and $\delta \mu_c$, with solution
\begin{eqnarray}
&&\delta T_c(t)^2 =  \frac{ L^2_{13}\Lambda_{13}- \Lambda_{11} L^2_{33}}{(\Lambda_{11}L_{22}^2-L^1_{22}L^2_{13}\hbar\Omega)} \;(\hbar\Omega)^2,\;\nonumber  \\\nonumber\\\nonumber\\ 
&&\delta \mu_c(t) = \frac{L^1_{22} L^2_{33}(\hbar\Omega)^2- L^2_{22}\Lambda_{13} \hbar \Omega}{(\Lambda_{11}L_{22}^2-L^1_{22}L^2_{13}\hbar\Omega)}.
\end{eqnarray}

From Eqs. (\ref{cheatto}) and (\ref{coeff1}), it is easy to notice that $L^2_{13}\Lambda_{13}- \Lambda_{11} L^2_{33}=0$, and then 
\beq
\delta T_c(t) =  0.
\edq
Moreover, the relation between $L_{13}^2$ and $L_{33}^2$ in Eq.~(\ref{cheatto}) and  $\Lambda_{13}=-\frac{\dot{V}}{\Gamma \Omega}\Lambda_{11}$ in (\ref{coeff1}), lead to the solution
\beq
\delta \mu_c(t) = \frac{\hbar}{\Gamma}\dot{V}+{\cal{O}}(\Omega^2).
\edq

Therefore, we find that Eq. (\ref{sol}) obtained for $k_BT\gg\hbar\Omega$ within linear response regime, remain valid at zero temperature for which higher order contributions take place. This is a strong result, that shows the universality of the behavior of $\delta T
_c$ and $\delta \mu_c$ for any temperature of the reservoir.

\end{document}